\documentclass[ALICE,manyauthors]{cernphprep}

\usepackage[comma,square,numbers,sort&compress]{natbib}
\usepackage{hyperref}
\usepackage{lineno}
\usepackage{graphicx,subfigure}

\begin{document}%

\begin{titlepage}
\PHyear{2016}
\PHnumber{018}       
\PHdate{26 January}  
%

\title{Anisotropic flow of charged particles\\in Pb--Pb collisions at $\mathbf{\sqrt{{\textit s}_{\mathbf NN}}}$=5.02 TeV}
\ShortTitle{Anisotropic flow at 5.02 TeV}   

\Collaboration{ALICE Collaboration\thanks{See Appendix~\ref{app:collab} for the list of collaboration members}}
\ShortAuthor{ALICE Collaboration} 

\begin{abstract}
We report the first results of elliptic ($v_2$), triangular ($v_3$) and quadrangular flow ($v_4$) of charged particles in Pb--Pb collisions at a center-of-mass energy per nucleon pair of $\sqrt{s_{_{\rm NN}}}=$ 5.02 TeV with the ALICE detector at the CERN Large Hadron Collider. The measurements are performed in the central pseudorapidity region $|\eta| <$ 0.8 and for the transverse momentum range 0.2 $< p_{\rm T} < $ 5 GeV/$c$. The anisotropic flow is measured using two-particle correlations with a pseudorapidity gap greater than one unit and with the multi-particle cumulant method. Compared to results from Pb--Pb collisions at $\sqrt{s_{_{\rm NN}}}=$ 2.76 TeV, the anisotropic flow coefficients $v_{2}$, $v_{3}$ and $v_{4}$ are found to increase by (3.0$\pm$0.6)\%, (4.3$\pm$1.4)\% and (10.2$\pm$3.8)\%, respectively, in the centrality range 0--50\%. This increase can be attributed mostly to an increase of the average transverse momentum between the two energies. The measurements are found to be compatible with hydrodynamic model calculations. This comparison provides a unique opportunity to test the validity of the hydrodynamic picture and the power to further discriminate between various possibilities for the temperature dependence of shear viscosity to entropy density ratio of the produced matter in heavy-ion collisions at the highest energies.
\end{abstract}
\end{titlepage}
\setcounter{page}{2}


The goal of studies with relativistic heavy-ion collisions is to investigate the Quark-Gluon Plasma (QGP), a state of matter where quarks and gluons move freely over distances large in comparison to the typical size of a hadron. The transition from normal nuclear matter to the QGP state is expected to occur at extreme values of energy density (0.2--0.5 GeV/fm$^3$, according to lattice Quantum Chromodynamics calculations~\cite{Borsanyi:2010cj,Bazavov:2014pvz}), which are accessible in ultra-relativistic heavy-ion collisions at the Large Hadron Collider (LHC)~\cite{Aamodt:2010cz,Adam:2015ptt}. The study of such collisions provides the unique opportunity to probe the properties of the QGP in a region of the QCD phase diagram where a cross-over between the deconfined phase and normal nuclear matter is expected~\cite{Odyniec:2013aaa,Adamczyk:2012ku,Adler:2013aqf,Gavai:2004sd,Braun-Munzinger:2015hba}.


Studies of the azimuthal anisotropy of particle production have contributed significantly to the characterization of the system created in heavy-ion collisions~\cite{Ollitrault:1992bk,Voloshin:2008dg}.
Anisotropic flow, which measures the momentum anisotropy of the final-state particles, is sensitive both to the initial geometry of the overlap region and to the transport properties and equation of state of the system. By using a general Fourier series decomposition of the azimuthal distribution of produced particles:
\begin{equation}
\frac{dN}{d\varphi} \propto 1 + 2\sum_{n=1}^{\infty} v_{n} \cos [n(\varphi-\Psi_{n})]\,,
\label{eq:Fourier}
\end{equation}
anisotropic flow is quantified with coefficients $v_n$ and corresponding symmetry planes $\Psi_{n}$~\cite{Voloshin:1994mz}. Due to the approximately ellipsoidal shape of the overlap region in non-central heavy-ion collisions (i.e.\ collisions that correspond to large impact parameter), the dominant flow coefficient is $v_2$, referred to as elliptic flow. In the transition from highest RHIC to LHC energies, elliptic flow increases by 30\%~\cite{Aamodt:2010pa}, as predicted by hydrodynamic models that include viscous corrections~\cite{Ackermann:2000tr, Luzum:2009sb,Hirano:2010jg,Hirano:2005xf,Drescher:2007uh}. 
Non-vanishing values of higher anisotropic flow harmonics $v_3$--$v_6$ at the LHC are ascribed primarily to the response of the produced QGP to fluctuations of the initial energy density profile of the colliding nucleons~\cite{Alver:2010gr, ALICE:2011ab, ATLAS:2012at, Chatrchyan:2013kba}. Moreover, due to such fluctuations each flow harmonic $v_n$ has a distinct symmetry plane $\Psi_n$ and recent measurements of their inter-correlations provide independent constraints on the QGP properties~\cite{Aad:2014fla}.
The combination of all of such results demonstrates that the shear viscosity to entropy density ratio ($\eta/s$) of the QGP produced in ultra-relativistic heavy-ion collisions at RHIC and LHC has a value close to $1/4\pi$, a lower bound obtained in strong-coupling calculations based on the AdS/CFT conjecture~\cite{Kovtun:2004de}.


Recently, predictions from Niemi et al. on anisotropic flow coefficients for Pb--Pb $\sqrt{s_{_{\rm NN}}}=5.02$ TeV collisions using the EKRT model were reported in~\cite{Niemi:2015voa}. These predictions have a special emphasis on the discriminating power between various parameterizations of the temperature dependence of $\eta/s$. It was argued that in the transition from 2.76 to 5.02 TeV, the elliptic flow estimated from two--particle correlations (denoted further in the text as $v_{2}\{2\}$, where the number in the curly brackets indicates the number of particles that are used in correlation~\cite{Borghini:2000sa}) can increase at most $\sim 5\%$ for all centrality classes. Details of the increase depend on the parameterization of $\eta/s(T)$. On the other hand, higher flow harmonic observables, like $v_{3}\{2\}$ and $v_{4}\{2\}$, are predicted to increase more rapidly,  $10-30\%$. With a different approach, where previously measured values of flow harmonics at lower LHC energies are taken as a baseline, Noronha-Hostler et al.~\cite{Noronha-Hostler:2015uye} predict a larger increase for both elliptic and triangular flow in peripheral compared to central collisions in transition from 2.76 to 5.02 TeV. They conclude that the anisotropic flow already reaches saturation and its maximum value in central collisions at 2.76 TeV.


A necessary condition for the development of anisotropic flow is the initial anisotropy in the interaction region of the two colliding ions. These coordinate space anisotropies are described in terms of eccentricities, that are not directly accessible experimentally. Nonetheless, the theoretical modeling of such eccentricities is actively being studied. For instance, hydrodynamic calculations based on a MC-Glauber model and MC-KLN initial conditions do not agree on the details of the saturation of elliptic flow at LHC energies~\cite{Shen:2012vn}. However, with these two initial state models, it was shown that the final spatial eccentricity decreases monotonically as the collision energy increases \cite{Shen:2012vn}, and is expected to become negative only at the very large collision energies available at the LHC (see Fig. 9 in \cite{Shen:2012vn}). 


In addition to the initial conditions, various other stages of evolution of the system in a heavy-ion collision may contribute to the development of anisotropic flow. At lower energies, the state of the system will resemble primarily a hadronic gas, and hadron rescattering is the dominant contribution to the anisotropic flow. At higher energies, anisotropic flow mostly develops in the thermalized color-deconfined QGP phase. However, even at these higher energies, the contribution from the hadronic phase can be significant. The relative amount of time the system spends in different phases varies with collision energy~\cite{Shen:2012vn,Auvinen:2013ira}. Radial flow, a measure for the average velocity of the system's collective radial expansion, also increases as a function of collision energy, which translates into more particles being transferred to a higher transverse momentum ($p_{\rm T}$) region, thus leading to an increase in average anisotropic flow values. On the other hand, the opposite dependence of differential $v_2(p_{\rm T})$ is expected for light (increase at low $p_{\rm T}$) and heavy particles (decrease at low $p_{\rm T}$) as a function of collision energy, which might yield to the saturation of elliptic flow signal~\cite{Shen:2012vn}. Finally, the relative importance of various stages in the system evolution as a function of collision energy can also vary for each flow coefficient~\cite{Auvinen:2013ira}.


The data used in this Letter were recorded with the ALICE detector~\cite{Aamodt:2008zz, Alessandro:2006yt} in November 2015 in Run 2 at the LHC with Pb--Pb collisions at $\sqrt{s_{_{\rm NN}}} =$ 5.02 TeV. Minimum bias Pb--Pb events were triggered by the coincidence of signals from the V0 detector. The V0 detector is composed of two arrays of scintillator counters, V0-A and V0-C, which cover the pseudorapidity ranges 2.8 $< \eta <$ 5.1 and -3.7 $< \eta <$ -1.7, respectively ~\cite{Aamodt:2008zz}. 
Centrality quantifies the fraction of geometrical cross-section of the colliding nuclei.
It is determined using the sum of the amplitudes of the V0-A and C signals, which provides a resolution better than 0.5$\%$ up to 20$\%$ central Pb--Pb collisions, and better than 2\% for peripheral collisions~\cite{Abelev:2013qoq}. The offline event selection employs the information from two Zero Degree Calorimeters (ZDC)~\cite{Aamodt:2008zz} positioned 112.5 m from the interaction point on either side. Beam background events are removed using timing information from the V0 and the ZDC, respectively. To ensure a uniform acceptance and reconstruction efficiency in the pseudorapidity region $|\eta|<$ 0.8, only events with a reconstructed vertex within 10 cm from the center of the detector along the beam direction were used. A sample of 140 k PbPb collisions events passed the selection criteria. Only one low luminosity run (with trigger rate of 27 Hz) was used, being least affected by pile-up and distortions from space charge in the main tracking detector, the Time Projection Chamber (TPC).


\begin{figure}[th!f]
\begin{center}
\includegraphics[width=0.6\textwidth]{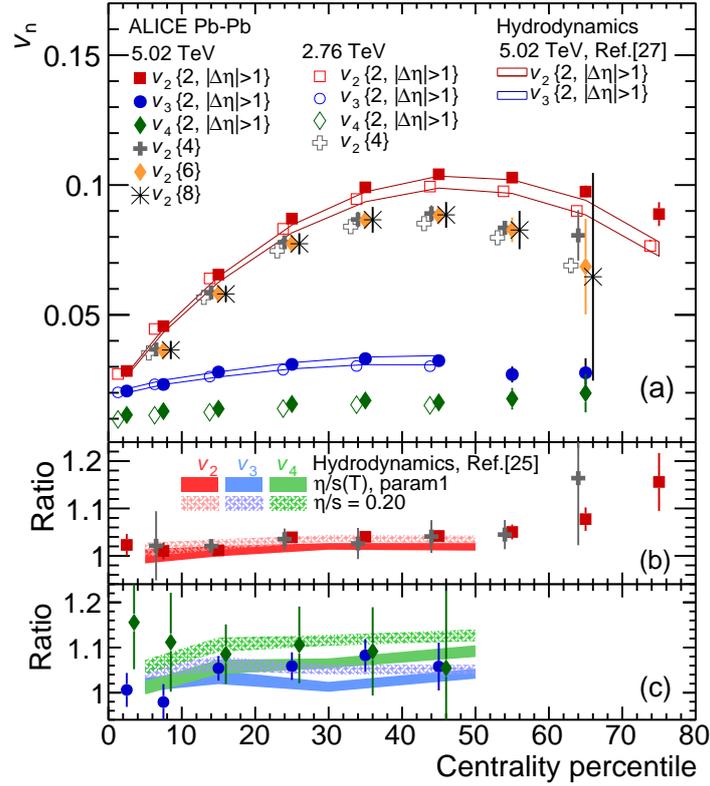}
\caption{(color online) (a) Anisotropic flow $v_{n}$ integrated over the $p_{\rm T}$ range 0.2 $< p_{\rm T} <$ 5.0 GeV/$c$, as a function of event centrality, for the two--particle (with $|\Delta\eta|>1$) and multi-particle cumulant methods. Measurements for Pb--Pb collisions at $\sqrt{s_{_{\rm NN}}}=$ 5.02 (2.76) TeV are shown by solid (open) markers ~\cite{ALICE:2011ab}. The ratios of $v_{2}\{2,|\Delta\eta|>1\}$ (red), $v_{2}\{4\}$ (gray) and $v_{3}\{2,|\Delta\eta|>1\}$ (blue), $v_{4}\{2,|\Delta\eta|>1\}$ (green) between Pb--Pb collisions at 5.02 TeV and 2.76 TeV, are presented in Fig.~\ref{fig1} (b) and (c). Various hydrodynamic calculations are also presented \cite{Noronha-Hostler:2015uye,Niemi:2015voa}. The statistical and systematical uncertainties are summed in quadrature (the systematic uncertainty is smaller than the statistical uncertainty, which is typically within 5\%). Data points are shifted for visibility.}
\label{fig1} 
\end{center}
\end{figure}
 
Charged tracks are reconstructed using the ALICE Inner Tracking System (ITS) and TPC~\cite{Aamodt:2008zz}. This combination ensures a high detection efficiency, optimum momentum resolution, and a minimum contribution from photon conversions and secondary charged particles produced either from the detector material or from weak decays. In order to reduce the contamination from secondary particles, only tracks with a distance of closest approach (DCA) to the interaction point of less than 3 cm, both in the longitudinal and transverse directions, are accepted. The tracking efficiency is calculated from a Monte Carlo simulation that uses HIJING \cite{Wang:1991hta} to simulate particle production. GEANT3~\cite{Brun:1994aa} is then used for transporting simulated particles, followed by a full calculation of the detector response (including production of secondary particles) and track reconstruction performed with the ALICE reconstruction framework. The tracking efficiency is $\sim 70\%$ at $p_{\rm T} \sim 0.2 $ GeV/$c$ and increases to an approximately constant value of $\sim 80\%$ for $p_{\rm T} > 1 $ GeV/$c$. The $p_{\rm T}$ resolution is better than 5\% for the region presented in this Letter. The systematic uncertainty related to the non-uniform reconstruction efficiency was found to be at the level of 1\%.
The flow coefficients from tracks that are reconstructed from TPC space points alone were compared to coefficients extracted from particles that used both TPC clusters and ITS hits, which were found to agree within $\sim$2\%. This difference was taken into account in the estimation of the systematic uncertainty. Altering the selection criteria for the tracks reconstructed with the TPC resulted in a variation of the results of 0.5\%, at most. Other selection criteria that have been scrutinized are the centrality determination, e.g. using the Silicon Pixel Detector (SPD), which contributed by less than 1\%, the polarity of the magnetic field of the ALICE detector and the position of the reconstructed primary vertex, whose contributions were found to be negligible. The systematic uncertainties evaluated for each of the sources mentioned above were added in quadrature to obtain the total systematic uncertainty of the measurements. 

\begin{figure}[th!f]
\begin{center}
\subfigure{\includegraphics[width=0.49\textwidth]{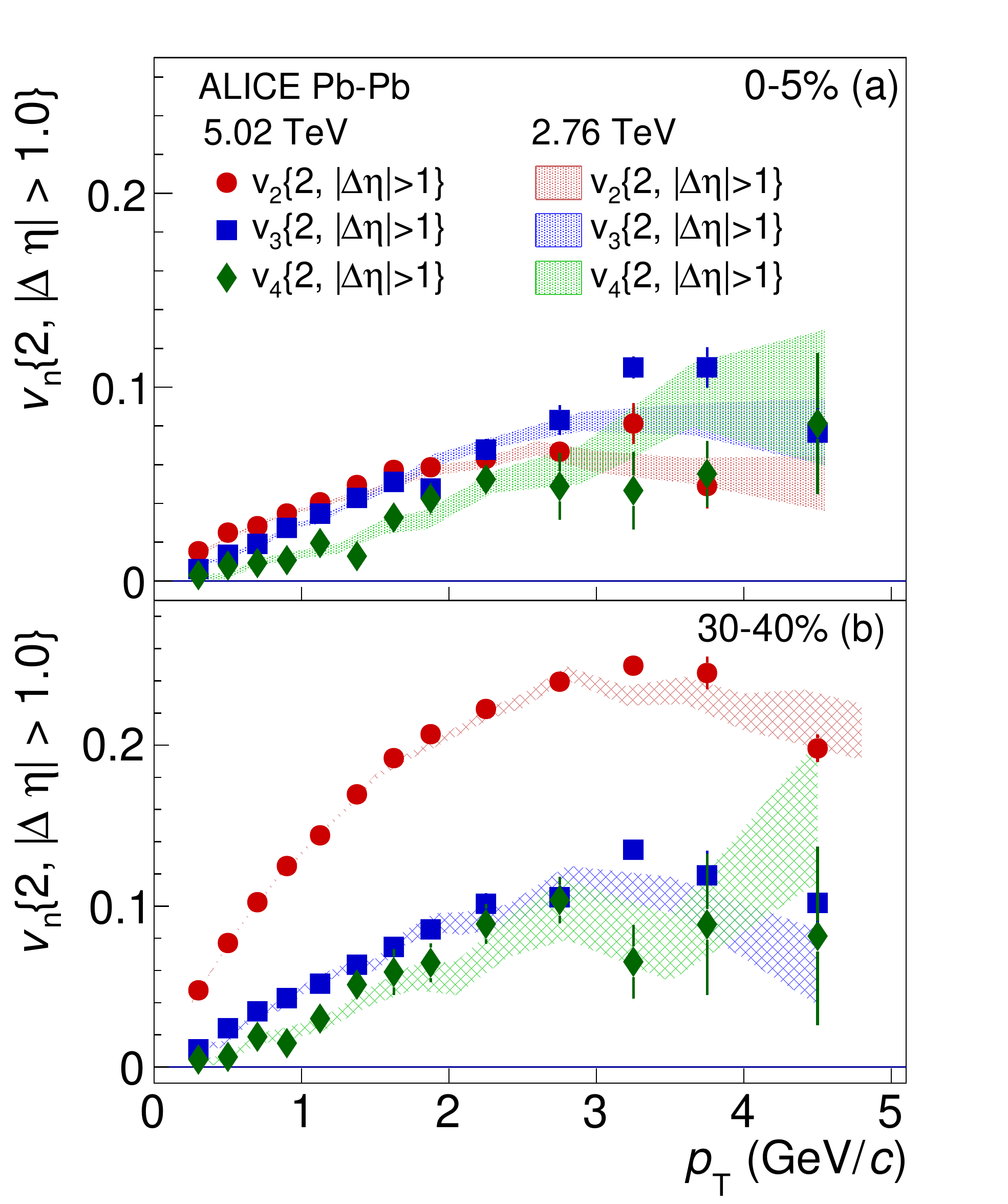}}
\subfigure{\includegraphics[width=0.49\textwidth]{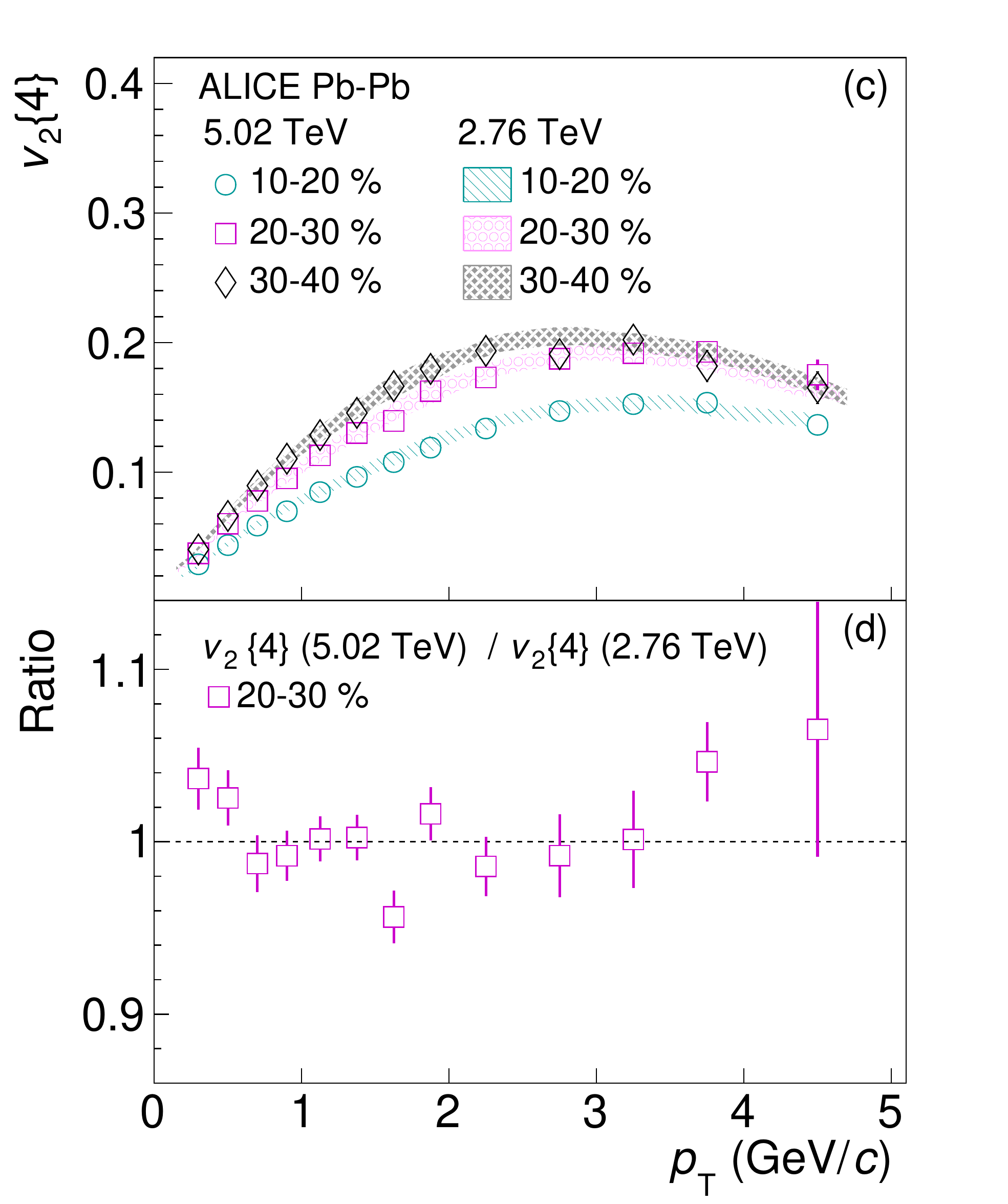}}
\caption{(color online) $v_{n}(p_{\rm T})$ using two--particle cumulant method with $|\Delta\eta|>$ 1 for (a) 0--5\% and (b) 30--40\% centrality classes;  (c) $v_{2}(p_{\rm T})$ using four-particle cumulant method for the centrality 10--20\%, 20--30\% and 30--40\%. Measurements for Pb--Pb collisions at $\sqrt{s_{_{\rm NN}}}=$ 2.76 TeV are also presented as shading. (d) The ratio of $v_{2}\{4\}$ in 20--30\% from two collision energies is also shown here. The statistical and systematical uncertainties are summed in quadrature (the systematic uncertainty is smaller than the statistical uncertainty, which is typically within 5\%).}.
\label{fig2} 
\end{center}
\end{figure}

In this Letter, we report the anisotropic flow measurements obtained from two- and multi-particle cumulants, using the approach proposed in~\cite{Zhou:2015iba, Bilandzic:2010jr, Bilandzic:2013kga}. These two measurements have different sensitivities to flow fluctuations and non-flow effects. Non-flow effects are azimuthal correlations not associated with the symmetry planes and usually arise from resonance decays and jets. Their contributions are expected to be suppressed when using a large pseudorapidity gap between particle pairs. Thus, in this study, we require a pseudorapidity gap of $|\Delta\eta| >$ 1. This observable is denoted as $v_{n}\{2,|\Delta\eta|>1\}$. On the other hand, non-flow contributions to multi-particle cumulants $v_{n}\{4\}$, $v_{n}\{6\}$, $v_{n}\{8\}$ are found to be negligible in events with large multiplicities characteristic for heavy-ion collisions~\cite{Abelev:2014mda}.


Figure~\ref{fig1} (a) presents the centrality dependence of $v_{2}$, $v_{3}$ and $v_{4}$ from two- and multi-particle cumulants, integrated over the $p_{\rm T}$ range 0.2 $< p_{\rm T} <$ 5.0 GeV/$c$, for 2.76 and 5.02 TeV Pb--Pb collisions. To elucidate the energy evolution of $v_2$, $v_3$ and $v_4$, the ratios of anisotropic flow measured at 5.02 TeV to 2.76 TeV are presented in Fig~\ref{fig1}.\ (b) and (c).
Assuming that non-flow effects are suppressed by the pseudorapidity gap, the remaining differences between two- and multi-particle cumulants of $v_{2}$ can be related to the strength of elliptic flow fluctuations, which are expected to give a positive and a negative contribution to the two- and multi-particle cumulant estimates, respectively~\cite{Voloshin:2008dg}. Moreover, the multi-particle cumulants $v_{2}\{4\}$, $v_{2}\{6\}$ and $v_{2}\{8\}$ are all observed to agree within 1\%, which indicates that non-flow effects are largely suppressed.
It is seen that $v_{2}\{2, |\Delta\eta|>1\}$ increases from central to peripheral collisions, and reaches a maximum value of 0.104 $\pm$ 0.001 (stat.) $\pm$ 0.002 (syst.) in the 40--50\% centrality class. For the higher harmonics, i.e.\ $v_3$ and $v_4$, the values are smaller and the centrality dependence is much weaker.

Furthermore, the predictions of anisotropic flow coefficients $v_n$ from the previously mentioned hydrodynamic model~\cite{Noronha-Hostler:2015uye} are compared to the measurements in Fig.~\ref{fig1}.\ (a). These predictions combine the changes in initial spatial anisotropy and the hydrodynamic response (treated as systematic uncertainty and shown by the width of the bands). The predictions are compatible with the measured anisotropic flow $v_n$ coefficients. At the same time, a different hydrodynamic calculation~\cite{Niemi:2015voa}, which employs both constant $\eta/s=$ 0.20 and temperature dependent $\eta/s$, can also describe the increase in anisotropic flow measurements of $v_{2}$ (shown in Fig~\ref{fig1}.\ (b)), $v_{3}$ and $v_{4}$ (see Fig~\ref{fig1}.\ (c)). In particular, among the different scenarios proposed in~\cite{Niemi:2015voa}, the measurements seem to favor a constant $\eta/s$ going from $\sqrt{s_{_{\rm NN}}} = $ 2.76 to 5.02~TeV Pb--Pb collisions. 

The increase of $v_{2}$ and $v_{3}$ from the two energies is rather moderate, while for $v_{4}$ it is more pronounced. In addition, none of the ratios of flow harmonics exhibit a significant centrality dependence in the centrality range 0--50\%, and thus the results of a fit with a constant value over these ratios are reported. An increase of (3.0$\pm$0.6)\%, (4.3$\pm$1.4)\% and (10.2$\pm$3.8)\%, is obtained for elliptic, triangular and quadrangular flow, respectively, over the centrality range 0--50\% in Pb--Pb collisions when going from 2.76 TeV to 5.02 TeV.
This increase of anisotropic flow is compatible with theoretical predictions described in~\cite{Niemi:2015voa, Noronha-Hostler:2015uye}. Overall, these measurements support a low value of $\eta/s$ for the system created in Pb--Pb collisions at $\sqrt{s_{_{\rm NN}}} = $ 5.02~TeV and seem to indicate that it does not increase significantly with respect to Pb--Pb collisions at $\sqrt{s_{_{\rm NN}}}=2.76$~TeV.

The anisotropic flow coefficients $v_{2}\{2, |\Delta\eta|>1\}$, $v_{3}\{2, |\Delta\eta|>1\}$ and $v_{4}\{2, |\Delta\eta|>1\}$ as a function of transverse momentum ($p_{\rm T}$) are presented in Fig.~\ref{fig2} for the 0--5\% and 30--40\% centrality classes. For the 0--5\% centrality class, at $p_{\rm T} >$ 2 GeV/$c$ $v_{3}\{2\}$ is observed to become larger than $v_{2}\{2\}$, while $v_{4}\{2\}$ is compatible with $v_{2}\{2\}$, within uncertainties. For the 30--40\% centrality class, we see that $v_{2}\{2\}$ is higher than $v_{3}\{2\}$ and $v_{4}\{2\}$ for the entire $p_{\rm T}$ range measured, with no crossing of the different order flow coefficients observed. 
Figure~\ref{fig2} (c) presents the $p_{\rm T}$-differential $v_{2}\{4\}$ for the 10--20\%, 20--30\% and 30--40\% centrality classes. The $v_{2}\{4\}$ decreases from mid-central to central collisions over the $p_{\rm T}$ range measured.
The comparison with the corresponding measurements from Pb--Pb collisions at $\sqrt{s_{_{\rm NN}}}=2.76$~TeV exhibits comparable values, as illustrated by the ratio of $v_{2}\{4\}$ for the two energies in Fig.~\ref{fig2} (d). This indicates that the increase observed in the $p_{\rm T}$ integrated flow results seen in Fig.~\ref{fig1} can be attributed to an increase of mean transverse momentum $\langle p_{\rm T} \rangle$. The measurements of $p_{\rm T}$-differential flow are more sensitive to initial conditions and $\eta/s$, and are expected to provide important information to constrain further details of the theoretical calculations, e.g. determination of radial flow and freeze-out conditions.

\begin{figure}[t]
\begin{center}
\includegraphics[width=0.7\textwidth]{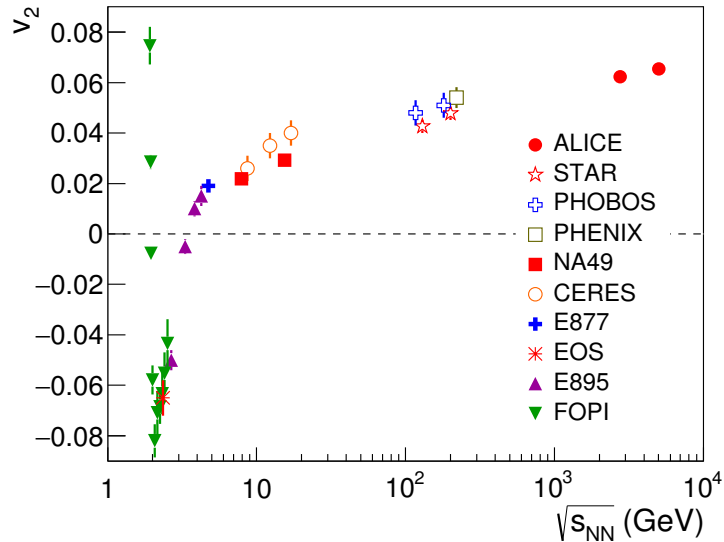}
\caption{Integrated elliptic flow ($v_{2}\{4\}$) for the 20--30\% most central Pb--Pb collisions at $\sqrt{s_{\rm NN}} = 5.02$ TeV compared with $v_{2}$ measurements at lower energies with similar centralities (see ~\cite{Aamodt:2010pa} for references to all data points).}
\label{fig3} 
\end{center}
\end{figure}

Figure~\ref{fig3} presents the comparison of the fully $p_{\rm T}$ integrated $v_2$ measured in the 20--30\% centrality in Pb--Pb collisions at the LHC with results at lower energies. This integrated value in the full $p_{\rm T}$ range is determined using two methods. The first uses fits to the efficiency-corrected charged-particle spectra and the $p_{\rm T}$-differential $v_{2}\{4\}$ presented in Fig.~\ref{fig2}, extrapolated to $p_{\rm T}=0$. The error on the integrated $v_2$ is estimated both from the uncertainty on the $p_{\rm T}$-differential measurements and from different parameterizations that provide a good fit of the data. The second calculates $v_{2}\{4\}$ using tracklets formed from SPD hits in the ITS, which have an acceptance of $p_{\rm T} \gtrsim$ 50 MeV/c. As each method uses different ALICE sub-detectors, they can provide independent measurements of $v_{2}$ coefficients. For this centrality range, they agree within 1\% for both energies. The values presented in the figure are weighted averages of these two measurements, using the inverse of the variance of each of them as weights. A continuous increase of anisotropic flow for this centrality has been observed from SPS/RHIC to LHC energies. For these fully $p_{\rm T}$ integrated coefficients, an increase of $4.9\pm1.9\%$ is observed going from $\sqrt{s_{_{\rm NN}}}=2.76$ to 5.02 TeV, which is close to values of the previously-mentioned hydrodynamic calculations \cite{Niemi:2015voa, Noronha-Hostler:2015uye}.

In summary, we have presented the first anisotropic flow measurements of charged particles in Pb--Pb collisions at $\sqrt{s_{_{\rm NN}}} =$  5.02 TeV at the LHC. An average increase of (3.0$\pm$0.6)\%, (4.3$\pm$1.4)\% and (10.2$\pm$3.8)\%, is observed for the transverse momentum integrated elliptic, triangular and quadrangular flow, respectively, over the centrality range 0--50\% going from 2.76 TeV to 5.02 TeV. The transverse momentum dependence of anisotropic flow has also been investigated, and does not change appreciably between the two LHC energies. Therefore, the increase in integrated flow coefficients can be attributed mostly to an increase in average transverse momentum. The measurements are found to be compatible with predictions from hydrodynamic models~\cite{Noronha-Hostler:2015uye,Niemi:2015voa}. Further comparisons of $p_{\rm T}$-differential flow measurements and theoretical calculations, which are not available at the time of writing, will provide extra constraints on the initial conditions and the transport properties of the QGP.

\newenvironment{acknowledgement}{\relax}{\relax}
\begin{acknowledgement}
\section*{Acknowledgements}

The ALICE Collaboration would like to thank all its engineers and technicians for their invaluable contributions to the construction of the experiment and the CERN accelerator teams for the outstanding performance of the LHC complex.
The ALICE Collaboration gratefully acknowledges the resources and support provided by all Grid centres and the Worldwide LHC Computing Grid (WLCG) collaboration.
The ALICE Collaboration acknowledges the following funding agencies for their support in building and
running the ALICE detector:
State Committee of Science,  World Federation of Scientists (WFS)
and Swiss Fonds Kidagan, Armenia;
Conselho Nacional de Desenvolvimento Cient\'{\i}fico e Tecnol\'{o}gico (CNPq), Financiadora de Estudos e Projetos (FINEP),
Funda\c{c}\~{a}o de Amparo \`{a} Pesquisa do Estado de S\~{a}o Paulo (FAPESP);
National Natural Science Foundation of China (NSFC), the Chinese Ministry of Education (CMOE)
and the Ministry of Science and Technology of China (MSTC);
Ministry of Education and Youth of the Czech Republic;
Danish Natural Science Research Council, the Carlsberg Foundation and the Danish National Research Foundation;
The European Research Council under the European Community's Seventh Framework Programme;
Helsinki Institute of Physics and the Academy of Finland;
French CNRS-IN2P3, the `Region Pays de Loire', `Region Alsace', `Region Auvergne' and CEA, France;
German Bundesministerium fur Bildung, Wissenschaft, Forschung und Technologie (BMBF) and the Helmholtz Association;
General Secretariat for Research and Technology, Ministry of Development, Greece;
National Research, Development and Innovation Office (NKFIH), Hungary;
Department of Atomic Energy and Department of Science and Technology of the Government of India;
Istituto Nazionale di Fisica Nucleare (INFN) and Centro Fermi -
Museo Storico della Fisica e Centro Studi e Ricerche ``Enrico Fermi'', Italy;
Japan Society for the Promotion of Science (JSPS) KAKENHI and MEXT, Japan;
Joint Institute for Nuclear Research, Dubna;
National Research Foundation of Korea (NRF);
Consejo Nacional de Cienca y Tecnologia (CONACYT), Direccion General de Asuntos del Personal Academico(DGAPA), M\'{e}xico, Amerique Latine Formation academique - 
European Commission~(ALFA-EC) and the EPLANET Program~(European Particle Physics Latin American Network);
Stichting voor Fundamenteel Onderzoek der Materie (FOM) and the Nederlandse Organisatie voor Wetenschappelijk Onderzoek (NWO), Netherlands;
Research Council of Norway (NFR);
National Science Centre, Poland;
Ministry of National Education/Institute for Atomic Physics and National Council of Scientific Research in Higher Education~(CNCSI-UEFISCDI), Romania;
Ministry of Education and Science of Russian Federation, Russian
Academy of Sciences, Russian Federal Agency of Atomic Energy,
Russian Federal Agency for Science and Innovations and The Russian
Foundation for Basic Research;
Ministry of Education of Slovakia;
Department of Science and Technology, South Africa;
Centro de Investigaciones Energeticas, Medioambientales y Tecnologicas (CIEMAT), E-Infrastructure shared between Europe and Latin America (EELA), 
Ministerio de Econom\'{i}a y Competitividad (MINECO) of Spain, Xunta de Galicia (Conseller\'{\i}a de Educaci\'{o}n),
Centro de Aplicaciones Tecnológicas y Desarrollo Nuclear (CEA\-DEN), Cubaenerg\'{\i}a, Cuba, and IAEA (International Atomic Energy Agency);
Swedish Research Council (VR) and Knut $\&$ Alice Wallenberg
Foundation (KAW);
Ukraine Ministry of Education and Science;
United Kingdom Science and Technology Facilities Council (STFC);
The United States Department of Energy, the United States National
Science Foundation, the State of Texas, and the State of Ohio;
Ministry of Science, Education and Sports of Croatia and  Unity through Knowledge Fund, Croatia;
Council of Scientific and Industrial Research (CSIR), New Delhi, India;
Pontificia Universidad Cat\'{o}lica del Per\'{u}.
\end{acknowledgement}

\bibliographystyle{utphys}   
\bibliography{bibliography}

\newpage
\appendix
\section{The ALICE Collaboration}
\label{app:collab}



\begingroup
\small
\begin{flushleft}
J.~Adam\Irefn{org39}\And
D.~Adamov\'{a}\Irefn{org84}\And
M.M.~Aggarwal\Irefn{org88}\And
G.~Aglieri Rinella\Irefn{org35}\And
M.~Agnello\Irefn{org110}\And
N.~Agrawal\Irefn{org47}\And
Z.~Ahammed\Irefn{org132}\And
S.~Ahmad\Irefn{org19}\And
S.U.~Ahn\Irefn{org68}\And
S.~Aiola\Irefn{org136}
% ...
\renewcommand\labelenumi{\textsuperscript{\theenumi}~}

\section*{Affiliation notes}
\renewcommand\theenumi{\roman{enumi}}
\begin{Authlist}
\item \Adef{0}Deceased
\item \Adef{idp1779728}{Also at: Georgia State University, Atlanta, Georgia, United States}
\item \Adef{idp3139616}{Also at: Also at Department of Applied Physics, Aligarh Muslim University, Aligarh, India}
\item \Adef{idp3840864}{Also at: M.V. Lomonosov Moscow State University, D.V. Skobeltsyn Institute of Nuclear, Physics, Moscow, Russia}
\end{Authlist}

\section*{Collaboration Institutes}
\renewcommand\theenumi{\arabic{enumi}~}
\begin{Authlist}

\item \Idef{org1}A.I. Alikhanyan National Science Laboratory (Yerevan Physics Institute) Foundation, Yerevan, Armenia
\item \Idef{org2}Benem\'{e}rita Universidad Aut\'{o}noma de Puebla, Puebla, Mexico
\item \Idef{org3}Bogolyubov Institute for Theoretical Physics, Kiev, Ukraine
\item \Idef{org4}Bose Institute, Department of Physics and Centre for Astroparticle Physics and Space Science (CAPSS), Kolkata, India
\item \Idef{org5}Budker Institute for Nuclear Physics, Novosibirsk, Russia
\item \Idef{org6}California Polytechnic State University, San Luis Obispo, California, United States
\item \Idef{org7}Central China Normal University, Wuhan, China
\item \Idef{org8}Centre de Calcul de l'IN2P3, Villeurbanne, France
\item \Idef{org9}Centro de Aplicaciones Tecnol\'{o}gicas y Desarrollo Nuclear (CEADEN), Havana, Cuba
\item \Idef{org10}Centro de Investigaciones Energ\'{e}ticas Medioambientales y Tecnol\'{o}gicas (CIEMAT), Madrid, Spain
\item \Idef{org11}Centro de Investigaci\'{o}n y de Estudios Avanzados (CINVESTAV), Mexico City and M\'{e}rida, Mexico
\item \Idef{org12}Centro Fermi - Museo Storico della Fisica e Centro Studi e Ricerche ``Enrico Fermi'', Rome, Italy
\item \Idef{org13}Chicago State University, Chicago, Illinois, USA
\item \Idef{org14}China Institute of Atomic Energy, Beijing, China
\item \Idef{org15}Commissariat \`{a} l'Energie Atomique, IRFU, Saclay, France
\item \Idef{org16}COMSATS Institute of Information Technology (CIIT), Islamabad, Pakistan
\item \Idef{org17}Departamento de F\'{\i}sica de Part\'{\i}culas and IGFAE, Universidad de Santiago de Compostela, Santiago de Compostela, Spain
\item \Idef{org18}Department of Physics and Technology, University of Bergen, Bergen, Norway
\item \Idef{org19}Department of Physics, Aligarh Muslim University, Aligarh, India
\item \Idef{org20}Department of Physics, Ohio State University, Columbus, Ohio, United States
\item \Idef{org21}Department of Physics, Sejong University, Seoul, South Korea
\item \Idef{org22}Department of Physics, University of Oslo, Oslo, Norway
\item \Idef{org23}Dipartimento di Fisica dell'Universit\`{a} 'La Sapienza' and Sezione INFN Rome, Italy
\item \Idef{org24}Dipartimento di Fisica dell'Universit\`{a} and Sezione INFN, Cagliari, Italy
\item \Idef{org25}Dipartimento di Fisica dell'Universit\`{a} and Sezione INFN, Trieste, Italy
\item \Idef{org26}Dipartimento di Fisica dell'Universit\`{a} and Sezione INFN, Turin, Italy
\item \Idef{org27}Dipartimento di Fisica e Astronomia dell'Universit\`{a} and Sezione INFN, Bologna, Italy
\item \Idef{org28}Dipartimento di Fisica e Astronomia dell'Universit\`{a} and Sezione INFN, Catania, Italy
\item \Idef{org29}Dipartimento di Fisica e Astronomia dell'Universit\`{a} and Sezione INFN, Padova, Italy
\item \Idef{org30}Dipartimento di Fisica `E.R.~Caianiello' dell'Universit\`{a} and Gruppo Collegato INFN, Salerno, Italy
\item \Idef{org31}Dipartimento di Scienze e Innovazione Tecnologica dell'Universit\`{a} del  Piemonte Orientale and Gruppo Collegato INFN, Alessandria, Italy
\item \Idef{org32}Dipartimento Interateneo di Fisica `M.~Merlin' and Sezione INFN, Bari, Italy
\item \Idef{org33}Division of Experimental High Energy Physics, University of Lund, Lund, Sweden
\item \Idef{org34}Eberhard Karls Universit\"{a}t T\"{u}bingen, T\"{u}bingen, Germany
\item \Idef{org35}European Organization for Nuclear Research (CERN), Geneva, Switzerland
\item \Idef{org36}Excellence Cluster Universe, Technische Universit\"{a}t M\"{u}nchen, Munich, Germany
\item \Idef{org37}Faculty of Engineering, Bergen University College, Bergen, Norway
\item \Idef{org38}Faculty of Mathematics, Physics and Informatics, Comenius University, Bratislava, Slovakia
\item \Idef{org39}Faculty of Nuclear Sciences and Physical Engineering, Czech Technical University in Prague, Prague, Czech Republic
\item \Idef{org40}Faculty of Science, P.J.~\v{S}af\'{a}rik University, Ko\v{s}ice, Slovakia
\item \Idef{org41}Faculty of Technology, Buskerud and Vestfold University College, Vestfold, Norway
\item \Idef{org42}Frankfurt Institute for Advanced Studies, Johann Wolfgang Goethe-Universit\"{a}t Frankfurt, Frankfurt, Germany
\item \Idef{org43}Gangneung-Wonju National University, Gangneung, South Korea
\item \Idef{org44}Gauhati University, Department of Physics, Guwahati, India
\item \Idef{org45}Helsinki Institute of Physics (HIP), Helsinki, Finland
\item \Idef{org46}Hiroshima University, Hiroshima, Japan
\item \Idef{org47}Indian Institute of Technology Bombay (IIT), Mumbai, India
\item \Idef{org48}Indian Institute of Technology Indore, Indore (IITI), India
\item \Idef{org49}Indonesian Institute of Sciences, Jakarta, Indonesia
\item \Idef{org50}Inha University, Incheon, South Korea
\item \Idef{org51}Institut de Physique Nucl\'eaire d'Orsay (IPNO), Universit\'e Paris-Sud, CNRS-IN2P3, Orsay, France
\item \Idef{org52}Institut f\"{u}r Informatik, Johann Wolfgang Goethe-Universit\"{a}t Frankfurt, Frankfurt, Germany
\item \Idef{org53}Institut f\"{u}r Kernphysik, Johann Wolfgang Goethe-Universit\"{a}t Frankfurt, Frankfurt, Germany
\item \Idef{org54}Institut f\"{u}r Kernphysik, Westf\"{a}lische Wilhelms-Universit\"{a}t M\"{u}nster, M\"{u}nster, Germany
\item \Idef{org55}Institut Pluridisciplinaire Hubert Curien (IPHC), Universit\'{e} de Strasbourg, CNRS-IN2P3, Strasbourg, France
\item \Idef{org56}Institute for Nuclear Research, Academy of Sciences, Moscow, Russia
\item \Idef{org57}Institute for Subatomic Physics of Utrecht University, Utrecht, Netherlands
\item \Idef{org58}Institute for Theoretical and Experimental Physics, Moscow, Russia
\item \Idef{org59}Institute of Experimental Physics, Slovak Academy of Sciences, Ko\v{s}ice, Slovakia
\item \Idef{org60}Institute of Physics, Academy of Sciences of the Czech Republic, Prague, Czech Republic
\item \Idef{org61}Institute of Physics, Bhubaneswar, India
\item \Idef{org62}Institute of Space Science (ISS), Bucharest, Romania
\item \Idef{org63}Instituto de Ciencias Nucleares, Universidad Nacional Aut\'{o}noma de M\'{e}xico, Mexico City, Mexico
\item \Idef{org64}Instituto de F\'{\i}sica, Universidad Nacional Aut\'{o}noma de M\'{e}xico, Mexico City, Mexico
\item \Idef{org65}iThemba LABS, National Research Foundation, Somerset West, South Africa
\item \Idef{org66}Joint Institute for Nuclear Research (JINR), Dubna, Russia
\item \Idef{org67}Konkuk University, Seoul, South Korea
\item \Idef{org68}Korea Institute of Science and Technology Information, Daejeon, South Korea
\item \Idef{org69}KTO Karatay University, Konya, Turkey
\item \Idef{org70}Laboratoire de Physique Corpusculaire (LPC), Clermont Universit\'{e}, Universit\'{e} Blaise Pascal, CNRS--IN2P3, Clermont-Ferrand, France
\item \Idef{org71}Laboratoire de Physique Subatomique et de Cosmologie, Universit\'{e} Grenoble-Alpes, CNRS-IN2P3, Grenoble, France
\item \Idef{org72}Laboratori Nazionali di Frascati, INFN, Frascati, Italy
\item \Idef{org73}Laboratori Nazionali di Legnaro, INFN, Legnaro, Italy
\item \Idef{org74}Lawrence Berkeley National Laboratory, Berkeley, California, United States
\item \Idef{org75}Moscow Engineering Physics Institute, Moscow, Russia
\item \Idef{org76}Nagasaki Institute of Applied Science, Nagasaki, Japan
\item \Idef{org77}National Centre for Nuclear Studies, Warsaw, Poland
\item \Idef{org78}National Institute for Physics and Nuclear Engineering, Bucharest, Romania
\item \Idef{org79}National Institute of Science Education and Research, Bhubaneswar, India
\item \Idef{org80}National Research Centre Kurchatov Institute, Moscow, Russia
\item \Idef{org81}Niels Bohr Institute, University of Copenhagen, Copenhagen, Denmark
\item \Idef{org82}Nikhef, Nationaal instituut voor subatomaire fysica, Amsterdam, Netherlands
\item \Idef{org83}Nuclear Physics Group, STFC Daresbury Laboratory, Daresbury, United Kingdom
\item \Idef{org84}Nuclear Physics Institute, Academy of Sciences of the Czech Republic, \v{R}e\v{z} u Prahy, Czech Republic
\item \Idef{org85}Oak Ridge National Laboratory, Oak Ridge, Tennessee, United States
\item \Idef{org86}Petersburg Nuclear Physics Institute, Gatchina, Russia
\item \Idef{org87}Physics Department, Creighton University, Omaha, Nebraska, United States
\item \Idef{org88}Physics Department, Panjab University, Chandigarh, India
\item \Idef{org89}Physics Department, University of Athens, Athens, Greece
\item \Idef{org90}Physics Department, University of Cape Town, Cape Town, South Africa
\item \Idef{org91}Physics Department, University of Jammu, Jammu, India
\item \Idef{org92}Physics Department, University of Rajasthan, Jaipur, India
\item \Idef{org93}Physik Department, Technische Universit\"{a}t M\"{u}nchen, Munich, Germany
\item \Idef{org94}Physikalisches Institut, Ruprecht-Karls-Universit\"{a}t Heidelberg, Heidelberg, Germany
\item \Idef{org95}Purdue University, West Lafayette, Indiana, United States
\item \Idef{org96}Pusan National University, Pusan, South Korea
\item \Idef{org97}Research Division and ExtreMe Matter Institute EMMI, GSI Helmholtzzentrum f\"ur Schwerionenforschung, Darmstadt, Germany
\item \Idef{org98}Rudjer Bo\v{s}kovi\'{c} Institute, Zagreb, Croatia
\item \Idef{org99}Russian Federal Nuclear Center (VNIIEF), Sarov, Russia
\item \Idef{org100}Saha Institute of Nuclear Physics, Kolkata, India
\item \Idef{org101}School of Physics and Astronomy, University of Birmingham, Birmingham, United Kingdom
\item \Idef{org102}Secci\'{o}n F\'{\i}sica, Departamento de Ciencias, Pontificia Universidad Cat\'{o}lica del Per\'{u}, Lima, Peru
\item \Idef{org103}Sezione INFN, Bari, Italy
\item \Idef{org104}Sezione INFN, Bologna, Italy
\item \Idef{org105}Sezione INFN, Cagliari, Italy
\item \Idef{org106}Sezione INFN, Catania, Italy
\item \Idef{org107}Sezione INFN, Padova, Italy
\item \Idef{org108}Sezione INFN, Rome, Italy
\item \Idef{org109}Sezione INFN, Trieste, Italy
\item \Idef{org110}Sezione INFN, Turin, Italy
\item \Idef{org111}SSC IHEP of NRC Kurchatov institute, Protvino, Russia
\item \Idef{org112}Stefan Meyer Institut f\"{u}r Subatomare Physik (SMI), Vienna, Austria
\item \Idef{org113}SUBATECH, Ecole des Mines de Nantes, Universit\'{e} de Nantes, CNRS-IN2P3, Nantes, France
\item \Idef{org114}Suranaree University of Technology, Nakhon Ratchasima, Thailand
\item \Idef{org115}Technical University of Ko\v{s}ice, Ko\v{s}ice, Slovakia
\item \Idef{org116}Technical University of Split FESB, Split, Croatia
\item \Idef{org117}The Henryk Niewodniczanski Institute of Nuclear Physics, Polish Academy of Sciences, Cracow, Poland
\item \Idef{org118}The University of Texas at Austin, Physics Department, Austin, Texas, USA
\item \Idef{org119}Universidad Aut\'{o}noma de Sinaloa, Culiac\'{a}n, Mexico
\item \Idef{org120}Universidade de S\~{a}o Paulo (USP), S\~{a}o Paulo, Brazil
\item \Idef{org121}Universidade Estadual de Campinas (UNICAMP), Campinas, Brazil
\item \Idef{org122}University of Houston, Houston, Texas, United States
\item \Idef{org123}University of Jyv\"{a}skyl\"{a}, Jyv\"{a}skyl\"{a}, Finland
\item \Idef{org124}University of Liverpool, Liverpool, United Kingdom
\item \Idef{org125}University of Tennessee, Knoxville, Tennessee, United States
\item \Idef{org126}University of the Witwatersrand, Johannesburg, South Africa
\item \Idef{org127}University of Tokyo, Tokyo, Japan
\item \Idef{org128}University of Tsukuba, Tsukuba, Japan
\item \Idef{org129}University of Zagreb, Zagreb, Croatia
\item \Idef{org130}Universit\'{e} de Lyon, Universit\'{e} Lyon 1, CNRS/IN2P3, IPN-Lyon, Villeurbanne, France
\item \Idef{org131}V.~Fock Institute for Physics, St. Petersburg State University, St. Petersburg, Russia
\item \Idef{org132}Variable Energy Cyclotron Centre, Kolkata, India
\item \Idef{org133}Warsaw University of Technology, Warsaw, Poland
\item \Idef{org134}Wayne State University, Detroit, Michigan, United States
\item \Idef{org135}Wigner Research Centre for Physics, Hungarian Academy of Sciences, Budapest, Hungary
\item \Idef{org136}Yale University, New Haven, Connecticut, United States
\item \Idef{org137}Yonsei University, Seoul, South Korea
\item \Idef{org138}Zentrum f\"{u}r Technologietransfer und Telekommunikation (ZTT), Fachhochschule Worms, Worms, Germany
\end{Authlist}
\endgroup

\end{document}